\pdfoutput=1 
\documentclass[aps,prd,twocolumn,superscriptaddress,amsfont,graphicx,nofootinbib,preprintnumbers]{revtex4}

\usepackage{color,graphicx,epsfig}
\usepackage{ifpdf}
\usepackage{amsmath}
\usepackage{bm}
\usepackage{color}
\usepackage[english]{babel}
\usepackage{graphicx}
\usepackage{amsfonts}
\usepackage{amssymb}
\usepackage{braket}
\usepackage{hyperref}
\usepackage{enumerate}

\definecolor{nicered}{rgb}{0.7,0.1,0.1}
\definecolor{nicegreen}{rgb}{0.1,0.5,0.1}
\hypersetup{colorlinks,citecolor= nicegreen,linkcolor= nicered}

\newcommand{\as}{\alpha_s}
\newcommand{\aso}{\bar\alpha_{s}}
\newcommand{\asb}{\alpha_{s,b}}
\newcommand{\asbo}{\bar\alpha_{s,b}}
\newcommand{\ep}{\epsilon}

\newcommand{\be}{\begin{equation}}
\newcommand{\ee}{\end{equation}}
\newcommand{\bea}{\begin{eqnarray}}
\newcommand{\eea}{\end{eqnarray}}

\definecolor{Red}{rgb}{1.,0.,0.}
\definecolor{randomcolour}{rgb}{0.2,0.5,0.7}

\DeclareMathAlphabet\mathbfcal{OMS}{cmsy}{b}{n}

\def\OMIT#1{}

\definecolor{darkred}{rgb}{0.9,0,0}

\definecolor{darkgreen}{rgb}{0,0,0.9}

\definecolor{darkblue}{rgb}{0,0,0.9}

\newcommand*{\Scale}[2][4]{\scalebox{#1}{$#2$}}

\allowdisplaybreaks[1]

\begin{document}
\def\OX{Rudolf Peierls Centre for Theoretical Physics, University of Oxford, Clarendon Laboratory, Parks
Road, Oxford OX1 3PU}
\def\TUM{Physik Department,
James-Franck-Straße 1, Technische Universit\"at M\"unchen,
D–85748 Garching, Germany}
\def\ORG{Exzellenzcluster ORIGINS, Boltzmannstr. 2, D-85748 Garching, Germany}
\def\MSU{Department of Physics and Astronomy, Michigan State University, East Lansing, Michigan 48824, USA}
\def\WDM{Wadham College, University of Oxford, Parks Road, Oxford OX1 3PN, UK}
\def\NEW{New College, University of Oxford, Holywell Street, Oxford OX1 3BN, UK}
\preprint{OUTP-21-28P, MSUHEP-21-035, TUM-HEP-1382/21}

\title{Three-loop gluon scattering in QCD and the gluon Regge trajectory}

\author{Fabrizio Caola}            
\email[Electronic address: ]{fabrizio.caola@physics.ox.ac.uk}
\affiliation{\OX} 
\affiliation{\WDM}

\author{Amlan Chakraborty}            
\email[Electronic address: ] {chakra69@msu.edu}
\affiliation{\MSU}

\author{Giulio Gambuti}            
\email[Electronic address: ]{giulio.gambuti@physics.ox.ac.uk}
\affiliation{\OX}
\affiliation{\NEW}

\author{Andreas von Manteuffel}            
\email[Electronic address: ]{vmante@msu.edu}
\affiliation{\MSU}

\author{Lorenzo Tancredi}            
\email[Electronic address: ]{lorenzo.tancredi@tum.de}
\affiliation{\TUM}
\affiliation{\ORG}

\begin{abstract}
We compute the three-loop helicity amplitudes for the scattering of four
gluons in QCD. We employ projectors in the 't
Hooft-Veltman scheme and construct the amplitudes from a minimal set of
physical building blocks,
which allows us to keep the computational
complexity under control. We obtain relatively compact results that
can be expressed in terms of harmonic polylogarithms. In addition,
we consider the Regge limit of our
amplitude and extract the gluon Regge
trajectory in full three-loop QCD.
This is the last missing ingredient required
for studying single-Reggeon exchanges
at next-to-next-to-leading logarithmic
accuracy. 
\end{abstract}

\maketitle
%%%%%
\section{Introduction}
Scattering amplitudes in Quantum Chromodynamics (QCD) are one of the
 fundamental ingredients to describe the dynamics of the high energy
 collision events produced at the Large Hadron Collider (LHC) at
 CERN. As a matter of fact, such probability amplitudes for processes
 involving four or five elementary particles and up to two loops in
 perturbation theory, are routinely used to measure the properties of
 Standard Model particles as the Higgs boson and to study its
 interactions with fermions and electroweak
 bosons~\cite{Heinrich:2020ybq}. Moreover, by providing the building
 blocks for precise estimates of Standard Model processes, they also
 allow us to put stringent constraints on New Physics signals
 predicted by various Beyond the Standard Model scenarios.

In addition to their practical use for collider physics, the analytic
calculation of scattering amplitudes in QCD provides an invaluable
source of information to understand general properties of perturbative
Quantum Field Theory (QFT).  In fact, with more loops and more
external particles participating to the scattering process, the
analytic structure of scattering amplitudes becomes increasingly rich,
in particular due to the appearance of new classes of special
functions, whose branch cut and analytical structure are to reproduce
those dictated by causality and unitarity in QFT.  In recent years, a
considerable effort has been devoted to study the properties of these
functions from first principles. The goal is to understand
whether an upper bound can be established for the type of mathematical
objects that can appear in the calculation of physically relevant
scattering processes.  While we are far from being able to provide a
complete answer to this question, the multitude of data collected in
the form of increasingly complicated amplitudes, have already revealed
crucial to discover and classify ubiquitous classes of such functions,
most notably the so-called multiple
polylogarithms~\cite{Remiddi:1999ew,Goncharov:2001iea,Vollinga:2004sn}
and more recently their elliptic
generalizations~\cite{BrownLevin,Bloch:2013tra,Adams:2015gva,Ablinger:2017bjx,Remiddi:2017har,Broedel:2017kkb}.
Most of these discoveries have been inspired by analytical results for
scattering amplitudes up to two loops, both in massless and in massive
theories, which have been an important focus of the efforts of the
particle physics community in the last two decades. A natural step
forward in these investigations is to push these calculations one loop
higher to determine which degree of generalization is required.
In combination with more general results on the simplified universal
properties of QCD in special kinematical limits,
perturbative calculations can also be used to have a glimpse of some
all-order QCD structures,
which only
emerge summing infinite classes of diagrams.  One of the classical
examples of such kinematical configurations 
is the so-called Regge
limit~\cite{Kuraev:1977fs}, where the energy of the colliding partons
is assumed to be much larger than the typical transferred momentum.
In this limit, the BFKL formalism~\cite{Kuraev:1976ge,Balitsky:1978ic}
allows one to reformulate the calculation of scattering amplitudes in
terms of the exchange of so-called reggeized gluons, which resum
specific contribution to the strong interaction among elementary
partons to all orders in the QCD coupling constants.

Motivated by these considerations, in this letter we focus on the
scattering of four gluons at three loops in QCD.  This is the most
complex of all scattering processes in QCD that involve four massless
particles, both for the number of terms involved in its calculation,
and also for its color and infrared structure.  As of today, this
process was known to three-loop order only in the simpler setting of $\mathcal N=4$ Super Yang
Mills (SYM) theory~\cite{Henn:2016jdu} and in the planar approximation
for pure Yang Mills theory~\cite{Jin:2019nya}.  The high-energy limit
of these results have been studied in
refs.~\cite{Falcioni:2021buo,DelDuca:2021vjq} respectively. In this
letter, we build upon the techniques that we have developed for the
calculations of simpler four-particle scattering
processes~\cite{Caola:2020dfu,Caola:2021rqz,Bargiela:2021wuy} and
compute the three loop scattering amplitudes for gluon-gluon
scattering in full, non-planar QCD.

We consider the process
\begin{equation}\label{eq:process}
g(p_1) + g(p_2)+ g(p_3) + g(p_4)  \rightarrow 0,
\end{equation}
where all momenta are taken to be incoming and massless
\begin{equation}\label{eq:mom_cons}
p_1^\mu + p_2^\mu  + p_3^\mu + p_4^\mu = 0, \quad p_i^2 = 0.
\end{equation}
The scattering process above can be parametrised in terms of  the usual set of Mandelstam invariants
\begin{align}\label{eq:mandelstams}
s\! =\! (p_1\! +\! p_2)^2, \; 
t  \! =\! (p_1 \!+ \!p_3)^2, \;
u \!=\! (p_2\! +\! p_3)^2,
\end{align}
which satisfy the relation $ u \!= \!- t \!-\!s $.  We work
in \textit{dimensional regularization} to regulate all ultraviolet and
infrared divergences.  More precisely, we adopt the 't Hooft-Veltman
scheme (tHV)~\cite{tHooft:1972tcz}, where loop momenta are taken to be
$d=4-2\ep$ dimensional, while momenta and polarizations associated
with external particles are kept in four dimensions.\\ The physical
scattering process $gg\rightarrow gg $ (relevant for di-jet
production) can be obtained from~\eqref{eq:process} by crossing
$p_{3,4} \! \rightarrow \! -p_{3,4}$.
In order to parametrize the kinematics for this process, it is useful
to define the dimensionless ratio
\begin{equation}
x = -t/s,
\end{equation}
so that in the physical
region $p_1+p_2 \to p_3+ p_4$ we have
\begin{equation} \label{eq:physical_region}
s>0, \: t<0, \: u<0; \quad 0<x<1.
\end{equation}

\section{Color and Lorentz Decomposition}
We write the scattering amplitude for 
$gg\to gg$ as
\begin{equation}\label{eq:physical_amplitude}
\mathcal{A}^{a_1 a_2 a_3 a_4} = 4 \pi \asb \, \sum_{i=1}^6. \mathcal{A}^{[i]} \mathcal{C}_i\, ,
\end{equation}
where $\asb$ is the bare strong
coupling, $\mathcal A^{[i]}$ are
color-ordered \emph{partial amplitudes},
and the color basis $\{\mathcal C_i\}$
reads
\begin{align}\label{eq:color_structures}
&\mathcal{C}_1 = \mathrm{Tr}[T^{a_1} T^{a_2} T^{a_3} T^{a_4}] +  \mathrm{Tr}[T^{a_1} T^{a_4} T^{a_3} T^{a_2}], \nonumber\\
&\mathcal{C}_2 = \mathrm{Tr}[T^{a_1} T^{a_2} T^{a_4} T^{a_3}]+\mathrm{Tr}[T^{a_1} T^{a_3} T^{a_4} T^{a_2}], \nonumber\\
&\mathcal{C}_3 = \mathrm{Tr}[T^{a_1} T^{a_3} T^{a_2} T^{a_4}] +  \mathrm{Tr}[T^{a_1} T^{a_4} T^{a_2} T^{a_3}],\nonumber\\
& \hspace{45pt} \mathcal{C}_4 =\mathrm{Tr}[T^{a_1} T^{a_2}]\mathrm{Tr}[T^{a_3} T^{a_4}], \nonumber\\
& \hspace{45pt} \mathcal{C}_5 = \mathrm{Tr}[T^{a_1} T^{a_3}]\mathrm{Tr}[T^{a_2} T^{a_4}], \nonumber\\
& \hspace{45pt} \mathcal{C}_6 = \mathrm{Tr}[T^{a_1} T^{a_4}]\mathrm{Tr}[T^{a_2} T^{a_3}].
\end{align}
Here the adjoint representation index $a_i$ corresponds to the $i$-th
external gluon, while $T^a_{ij}$ are the fundamental $SU(N_c)$
generators normalised such that $\mathrm{Tr}[T^aT^b]
= \frac{1}{2} \delta^{ab}$.  As it is well known, the partial
amplitudes $ \mathcal{A}^{[i]}$ are independently gauge invariant. The
advantage of using a color-ordered decomposition is that, by
construction, the amplitudes $ {\mathcal{A}}^{[i]}$ are not all
independent under crossings of the external momenta.  We can restrict
ourselves to compute only two of the structures above and obtain all
other partial amplitudes by crossing symmetry. For definiteness, we
choose to focus on $ \mathcal{A}^{[1]}$ and
$\mathcal{A}^{[4]}$.

In order to compute $ \mathcal{A}^{[1]}$ and $\mathcal{A}^{[4]}$, it is convenient to   
 further decompose them with respect to a basis of Lorentz covariant
tensor structures.  In the following we denote the polarization vector of the 
$i$-th external gluon as $\epsilon(p_i) = \epsilon_i$, which satisfies the transversality condition 
$\epsilon_i \! \cdot \! p_i = 0$.  By making the cyclic gauge choice $\epsilon_i  \! \cdot \! p_{i+1} = 0$, with $p_5 = p_1$,
and restricting ourselves to physical four-dimensional external states, 
one finds \cite{Peraro:2019cjj,Peraro:2020sfm} that each partial amplitude can be decomposed 
as
\begin{equation}\label{eq:tensor_decomp}
\mathcal{A}^{[j]}(s,t) = \sum_{i=1}^{8} \mathcal{F}^{[j]}_i \: T_i,
\end{equation}
where the coefficient functions $\mathcal{F}^{[j]}_i$ are usually referred to as \textit{form factors}
and the tensors $T_i$ are defined as
\begin{align} \label{eq:Tensors}
&T_1 = \epsilon_1 \! \cdot \! p_3\; \epsilon_2 \! \cdot \! p_1\; \epsilon_3 \! \cdot \! p_1\; \epsilon_4 \! \cdot \! p_2 \;, \nonumber \\
&T_2 = \epsilon_1 \! \cdot \! p_3\; \epsilon_2 \! \cdot \! p_1\; \epsilon_3 \! \cdot \! \epsilon_4 , \quad
T_3 = \epsilon_1 \! \cdot \! p_3\; \epsilon_3 \! \cdot \! p_1\; \epsilon_2 \! \cdot \! \epsilon_4 , \nonumber \\
&T_4 = \epsilon_1 \! \cdot \! p_3\; \epsilon_4 \! \cdot \! p_2\; \epsilon_2 \! \cdot \! \epsilon_3, \quad T_5 = \epsilon_2 \! \cdot \! p_1\; \epsilon_3 \! \cdot \! p_1\; \epsilon_1 \! \cdot \! \epsilon_4 , \nonumber \\
&T_6 = \epsilon_2 \! \cdot \! p_1\; \epsilon_4 \! \cdot \! p_2\; \epsilon_1 \! \cdot \! \epsilon_3 , \quad
T_7 = \epsilon_3 \! \cdot \! p_1\; \epsilon_4 \! \cdot \! p_2\; \epsilon_1 \! \cdot \! \epsilon_2 , \nonumber \\
&T_8 = \epsilon_1 \! \cdot \! \epsilon_2\;  \epsilon_3 \! \cdot \! \epsilon_4+ \epsilon_1 \! \cdot \! \epsilon_4\;  \epsilon_2 \! \cdot \! \epsilon_3 + \epsilon_1 \! \cdot \! \epsilon_3\;  \epsilon_2 \! \cdot \! \epsilon_4 \;.
\end{align}
The form factors can be extracted by defining a set 
of eight projectors $P_i$ which are in one to one correspondence with the 
tensors in eq.~\eqref{eq:Tensors},
such that $P_i \cdot T_j =\sum_{pol} P_i T_j= \delta_{ij}$.  

\section{Helicity Amplitudes}
In this letter we are ultimately
interested in the helicity amplitudes
$\mathcal{A}_{\bm \lambda}$, where ${\bm\lambda} = \{\lambda_1, \lambda_2, \lambda_3, \lambda_4\}$ and $\lambda_i$ is
the helicity of the $i$-th external particle.
In the four-gluon case we need to consider $2^4=16$
possible helicity choices. 
 However, only 8 helicity amplitudes are independent as
the remaining ones can be obtained by parity
conjugation, which effectively transforms the helicities as
${\bm\lambda} \rightarrow -{\bm\lambda}$.
The independent helicity amplitudes are in one to one correspondence
with the Lorentz tensors of eq.~\eqref{eq:Tensors} and their color
stripped counterparts can in fact be written as a linear combination
of the form factors $\mathcal{F}^{[j]}_i$.
In order to make this relation explicit, we start from the tensor decomposition 
in eq.~\eqref{eq:tensor_decomp} and employ the
 \textit{spinor-helicity formalism}~\cite{Dixon:1996wi} to fix the helicities of the external gluons.
We write the gluon polarization vectors for fixed $\pm$ helicity as 
\begin{equation}\label{eq:polvec}
\epsilon_{i,+}^\mu = \frac{[i+1|\gamma^\mu|i\rangle}{\sqrt{2} [i|i+1]}, \quad\quad \epsilon_{i,-}^\mu = \frac{[i|\gamma^\mu|i+1\rangle}{\sqrt{2} \langle i+1|i\rangle},
\end{equation}
where we used the cyclic gauge choice introduced above, identifying $|5] \equiv |1]$ and $|5\rangle \equiv |1\rangle$.
By inserting the specific representation of eq.~\eqref{eq:polvec} in eq.~\eqref{eq:tensor_decomp}, we can
write the color-ordered partial amplitudes as
\begin{equation}\label{eq:spinor_factorisation}
\mathcal{A}^{[i]}_{\bm{\lambda}}  = \mathcal{H}^{[i]}_{\bm{\lambda}}   \; s_{\bm{\lambda} },
\end{equation}
where  $s_{\bm\lambda}$ is a phase that carries all the spinor weight.
The decomposition~\eqref{eq:spinor_factorisation} is not unique. Here we follow~\cite{Bern:2001df}
and choose
\begin{align}\label{eq:spinor_factors}
&s_{++++}= \frac{\langle12\rangle\langle34\rangle}{ [12][34]}, \quad \quad \hspace{8pt}  
s_{-+++}= \frac{ \langle34\rangle\langle23\rangle\langle24\rangle}{\langle12\rangle\langle14\rangle[24]}, \nonumber \\
&s_{+-++} =  \frac{\langle34\rangle\langle13\rangle\langle14\rangle}{\langle21\rangle\langle24\rangle[14]}, \quad  
s_{++-+} =  \frac{\langle14\rangle\langle21\rangle\langle24\rangle}{ \langle32\rangle\langle34\rangle[24]}, \nonumber \\
&s_{+++-} =  \frac{\langle13\rangle\langle23\rangle\langle12\rangle}{ \langle42\rangle\langle14\rangle[12]}, \quad 
s_{++--} = \frac{ \langle12\rangle[34] }{[12]\langle34\rangle }, \nonumber \\
&s_{+-+-} =  \frac{ \langle13\rangle[24] }{ [13]\langle24\rangle } , \quad \quad \hspace{10pt} 
s_{+--+}=  \frac{ \langle14\rangle[23] }{ [14]\langle23\rangle } .
\end{align}

From now on we will focus on the calculation of $\mathcal{H}^{[j]}_{\bm{\lambda}}$,  which we will refer to as helicity amplitudes, with a slight abuse of notation.
The $\mathcal{H}^{[j]}_{\bm{\lambda}}$ can be expanded in terms of the bare QCD coupling in the usual way:
\begin{align}\label{eq:divergent_espansion}
\mathcal{H}_{\bm{\lambda}} &= \sum_{k=0}^3 \asbo^k S^k_\epsilon \mathcal{H}_{\bm{\lambda}}^{(k)} + \mathcal{O}(\asbo^4),
\end{align}
where we have omitted the color structure index $[j]$ for ease of reading 
and defined $\asbo = {\asb}/({4\pi})$ and $S_\ep = (4 \pi)^\ep e^{- \ep \gamma_E}$.
Here we focus on the computation of the three-loop amplitude $\mathcal{H}_{\bm{\lambda}}^{(3)}$.  As a byproduct 
we also re-computed the tree level, one- and two-loop amplitudes as a check of our framework
and
 found prefect agreement with previous results in the literature~\cite{Glover:2001af, Ahmed:2019qtg} . 
 
We use \texttt{QGRAF}~\cite{Nogueira:1991ex} to produce the relevant Feynman diagrams: there are 4 different
diagrams at tree level,  81 at one loop,  1771 at two loops and 48723 at three loops.  
We then use \texttt{FORM}~\cite{Vermaseren:2000nd} to apply the projection operators $P_i$ to 
suitable combinations of the Feynman diagrams and in this way write 
the helicity amplitudes  $\mathcal{H}^{[1]}_{\bm{\lambda}}$, 
$\mathcal{H}^{[4]}_{\bm{\lambda}}$
as linear combination of scalar Feynman integrals.
The integrals appearing in the computation of these amplitudes can be written as\\

\begin{equation}\label{eq:integrals}
\mathcal{I}^\text{top}_{n_1,...,n_N} = \mu_0^{2L\epsilon} e^{L \epsilon \gamma_E}  \int \prod_{i=1}^L \left( \frac{\mathrm{d}^d k_i}{i \pi^{\frac{d}{2}}} \right) \frac{1}{D_1^{n_1} \dots D_N^{n_N}}
\end{equation}

where $L$ stands for the number of loops, $k_i$ are the loop momenta,
$\gamma_E \approx 0.5772$ is the Euler constant, $\mu_0$ is the
dimensional regularization scale and $\ep = (4-d)/2$ is the
dimensional regulator.  Here ``top" can be any of the planar or
non-planar integral families which are given explicitly in
ref.~\cite{Caola:2021rqz}.  At three loops we find that a staggering
number of $ \sim \mathcal{O}(10^7)$ scalar
integrals contribute to the amplitude.  However, these integrals
are not linearly independent and can be related using symmetry relations and
integration by parts
identities~\cite{Chetyrkin:1981qh,Laporta:2001dd}. We performed this reduction
with \texttt{Reduze\;2}~\cite{Studerus:2009ye,vonManteuffel:2012np}
and \texttt{Finred}, an in-house implementation based on
Laporta's algorithm, finite field techniques~\cite{vonManteuffel:2014ixa,vonManteuffel:2016xki,Peraro:2016wsq,Peraro:2019svx}
and syzygy algorithms~\cite{Gluza:2010ws,Schabinger:2011dz,Ita:2015tya,Larsen:2015ped,Boehm:2017wjc,Agarwal:2019rag}.
In this way we were able to express the helicity amplitudes in terms
of the 486 \textit{master integrals} (MIs), which were first computed
in ref.~\cite{Henn:2020lye} and more recently in
ref.~\cite{Bargiela:2021wuy} in terms of simple harmonic
polylogarithms (HPLs)~\cite{Remiddi:1999ew}.  After inserting the
analytic expressions for the master integrals, we obtain
the \emph{bare} helicity amplitudes
$\mathcal{H}^{(j)}_{\bm \lambda}$ as a Laurent series in $\ep$ up
to $\mathcal O(\ep^0)$ in terms of HPLs up to transcendental weight
six.

\section{UV renormalization and IR behavior}
The bare helicity amplitudes 
contain both ultraviolet (UV) and infrared (IR) divergencies that manifest as poles in the series expansions of the dimensional regulator $\ep$.
UV divergences can be removed by expressing the amplitudes in terms of  the 
$\overline{\text{MS}}$ renormalized strong coupling $\as(\mu)$ using
\begin{equation}\label{eq:ren_coupling}
\asbo \mu_0^{2\ep} S_\ep = \aso(\mu)  \mu^{2\ep} Z\left[\aso(\mu) \right],
\end{equation}
where $\aso(\mu) = \as(\mu)/(4\pi)$,
$\mu$ is the renormalization scale
and
\begin{align}\label{eq:Zuv}
Z[\aso]  &=  1
- \aso  \frac{ \beta_0 }{\epsilon } +\aso^2 \left( \frac{\beta_0^2}{\epsilon^2} - \frac{\beta_1 }{2 \epsilon} \right)  \nonumber \\
&\quad
-\aso^3  \left( \frac{\beta_0^3}{\epsilon^3} - \frac{ 7}{6} \frac{\beta_0 \beta_1}{\epsilon^2}+ \frac{\beta_2}{3 \epsilon} \right)  + \mathcal{O}(\aso^4)\,.
\end{align}
The explicit form of the $\beta$-function coefficients $\beta_i$ is immaterial for our discussion; for the reader's convenience,
we provide them in the Supplemental Material.
The UV-renormalized helicity amplitudes $\mathcal{H}_{{\bm{\lambda}},\: \text{ren}}$
are obtained by expanding eq.~\eqref{eq:physical_amplitude} in $\aso(\mu)$. In particular, $\mathcal{H}^{(k)}_{{\bm{\lambda}},\: \text{ren}}$ is the
(color- and helicity-stripped)
coefficient of the $\aso^k$ term. 

The renormalized amplitudes still contain poles of IR origin,
whose structure is universal.
The infrared structure of QCD  scattering amplitudes was first 
studied at two loops in~\cite{Catani:1998bh} and later extended to general processes and three loops in~\cite{Sterman:2002qn,Aybat:2006wq,Aybat:2006mz,Becher:2009cu,Becher:2009qa,Dixon:2009gx,Gardi:2009qi,Gardi:2009zv,Almelid:2015jia}.
Up to three loop order, one can write~\cite{Becher:2009cu,Becher:2009qa}
\begin{equation}\label{eq:IR_factorisation}
\mathcal{H}_{{\bm{\lambda}},\: \text{ren}} = \mathbfcal{Z}_{IR} \; \mathcal{H}_{{\bm{\lambda}},\: \text{fin}} \; ,
\end{equation}
where $ \mathcal{H}_{{\bm{\lambda}},\: \text{fin}} $ are \emph{finite
remainders} and $\mathbfcal Z_{IR}$ is a color matrix that acts on the
$\{\mathcal C_i\}$ basis ~\eqref{eq:color_structures}.  It can be
written in terms of the so-called soft anomalous dimension
$\mathbf\Gamma$ as
\begin{equation}\label{eq:exponentiation}
\mathbfcal{Z}_{IR} = \mathbb{P}\exp \left[ \int_\mu^\infty \frac{\mathrm{d} \mu'}{\mu'}  \mathbf{\Gamma}(\{p\},\mu')\right]  \; ,
\end{equation}
where the \textit{path ordering} operator $\mathbb{P}$ reorganizes
color operators in increasing values of $\mu'$ from left to
right and is immaterial up to three loops
since to this order $[\mathbf\Gamma(\mu),\mathbf\Gamma(\mu')] = 0$.
The soft
anomalous dimension can be written as
\begin{equation}\label{eq:dipole_+_quadrupole}
\mathbf{\Gamma}=  \mathbf{\Gamma}_{\text{dip}}  + \mathbf{\Delta}_4  \; .
\end{equation}
The \emph{dipole} term $\mathbf{\Gamma}_{\text{dip}}$ is due to the pairwise exchange of color charge between external legs and reads
\begin{align}\label{eq:dipole}
\mathbf{\Gamma}_{\text{dip}}  &=  \sum_{1\leq i < j \leq 4} \mathbf{T}^a_i \; \mathbf{T}^a_j\; \gamma^\text{K} \; \ln\Scale[1.15]{\left(\frac{\mu^2}{-s_{ij}-i \delta}\right)}  + 4 \gamma^g \; ,
\end{align}
where $s_{ij} = 2p_i\cdot p_j$, $\gamma^{\text{K}}$ is the \textit{cusp anomalous dimension}
\cite{Korchemsky:1987wg,Moch:2004pa,Vogt:2004mw,Grozin:2014hna,Henn:2019swt,Huber:2019fxe,vonManteuffel:2020vjv} and $\gamma^g$ is the \textit{gluon anomalous dimension} \cite{Ravindran:2004mb,Moch:2005id,Moch:2005tm,Agarwal:2021zft}.
Their explicit form up to the order $\aso^3$ required here is reproduced in the Supplemental Material for convenience.
In eq.~\eqref{eq:dipole} we have also introduced the standard color insertion operators $\mathbf{T}^a_i$, which only act
on the $i$-th external color index. In particular, in our case their action on $\{\mathcal C_i\}$ is defined as
$\mathbf{T}_i^a T^{b_i} = -i f^{a b_i c_i} T^{c_i} = [T^{b_i},T^a]$.

The \emph{quadrupole} contribution ${\bm \Delta}_4$ in
eq.~\eqref{eq:dipole_+_quadrupole} accounts instead for the exchange
of color charge among (up to) four external legs. It becomes relevant for
the first time at three loops, $ \mathbf{\Delta}_4
= \sum_{n=3}^\infty \aso^n \mathbf{\Delta}^{(n)}_4 $, where it
reads~\cite{Almelid:2015jia}
\begin{align} \label{eq:quadrupole}
&\mathbf{\Delta}^{(3)}_4 = f_{abe} f_{cde}\bigg[
- 16 \,C   \, \sum_{i=1}^4 \sum_{\substack{1\leq j < k \leq4 \\ j,k\neq i}}  \left\{ \mathbf{T}^a_i,\mathbf{T}^d_i \right\} \mathbf{T}^b_j \mathbf{T}^c_k 
\nonumber \\& + 
128 \left[ \mathbf{T}^a_1\mathbf{T}^c_2\mathbf{T}^b_3\mathbf{T}^d_4D_1(x) - \mathbf{T}^a_4\mathbf{T}^b_1\mathbf{T}^c_2\mathbf{T}^d_3 D_2(x) \right]
\bigg],
\end{align}
with  $C = \zeta_5 + 2 \zeta_2 \zeta_3$. The functions $D_1(x)$ and $D_2(x)$ in our
notation are  reported in the Supplemental Material.

\begin{figure}[th]
    \includegraphics[width=1.\linewidth]{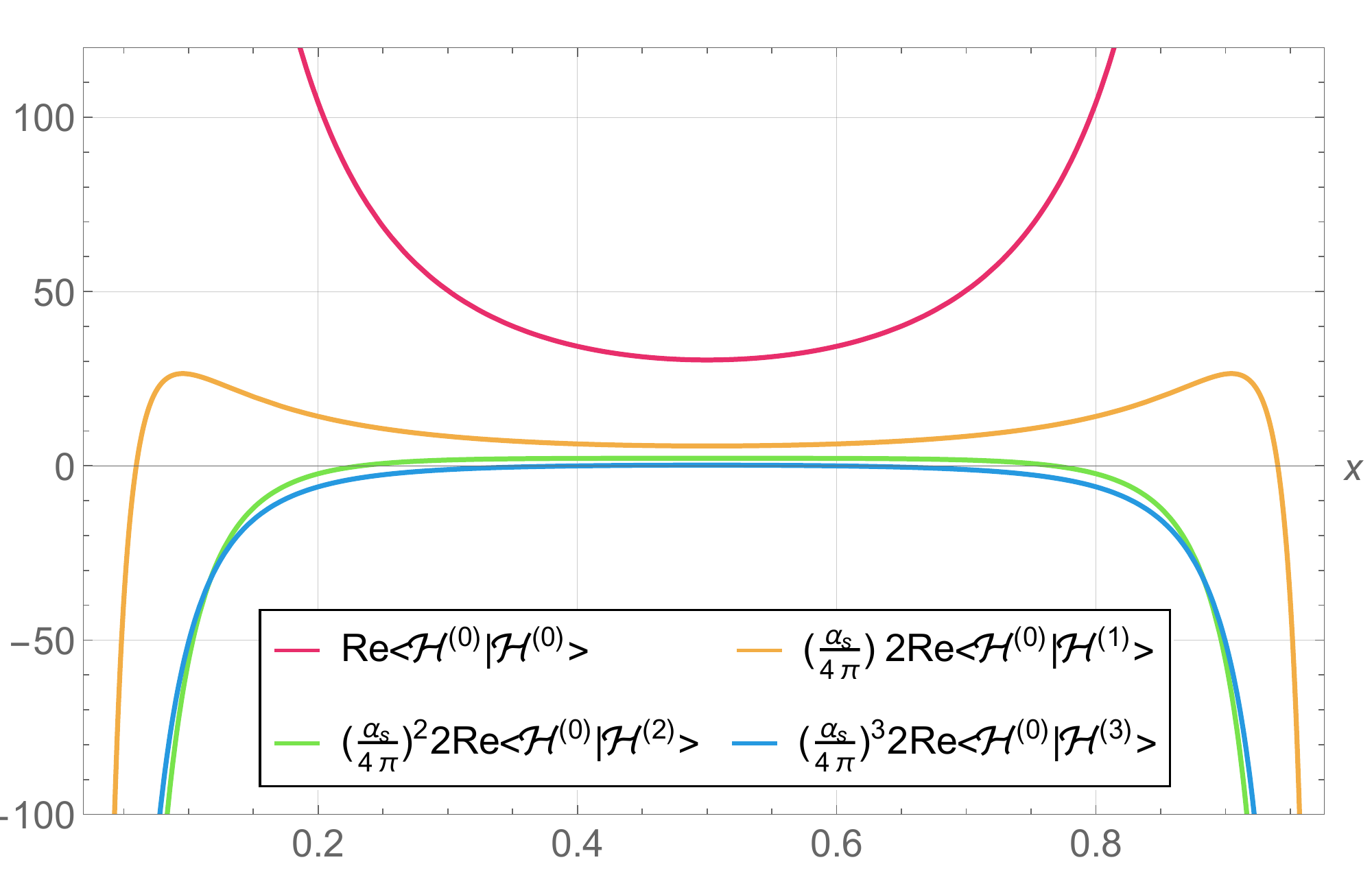} \caption{Tree level amplitude squared and interferences of tree level with $L=1,2,3$ loop amplitudes in dependence of $x=-t/s$.
}  \label{fig:helamp}
\end{figure}

We verified that the IR singularities of our three-loop amplitudes match perfectly those generated by 
eqs.~\eqref{eq:IR_factorisation}-\eqref{eq:quadrupole},
which provides a highly non-trivial check of our results. Our results for the finite
remainder $ \mathcal{H}_{{\bm{\lambda}},\: \text{fin}} $ are relatively compact, but
still 
too long to be presented here. They 
are included in computer-readable format
in the ancillary files accompanying the
\texttt{arXiv} submission of this manuscript.
In fig.~\ref{fig:helamp}, we plot our results for 
the interference with the tree level, defined as
\begin{align}
\langle \mathcal H^{(0)} | 
\mathcal H^{(L)}\rangle 
 \equiv \mathcal N
 \sum_{i,j=1}^6
 \mathcal C_i^\dagger \mathcal C_j
 \sum_{\bm \lambda}
 \mathcal H^{[i],(0)^*}_{\bm\lambda}
 \mathcal H^{[j],(L)}_{\bm\lambda, \rm fin},
\end{align}
where $\mathcal N = 1/[ 2(N_c^2-1)]^2$
is the initial-state color and helicity
averaging factor and the polarization sum
runs over all the 16 helicity configurations. 
Further, we have set $\mu^2=s$, $\alpha_s=0.118$, $N_c=3$ and $n_f=5$.
  
\section{High Energy Limit and the Gluon Regge trajectory}
QFT scattering amplitudes exhibit interesting factorisation properties in the high energy (Regge) limit.  
In terms of the variables introduced in this letter,  this limit corresponds to $|s| \approx |u| \gg |t|  $, or equivalently $x \rightarrow 0 $.
For studying this region it is convenient to split scattering amplitudes 
into parts of definite signature under the $s \leftrightarrow u$ exchange:
\begin{equation}
  \mathcal{H}_{\mathrm{ren,\pm}} = \frac{1}{2}\left.[ \mathcal{H}_{\mathrm{ren}}(s,u)   \pm   \mathcal{H}_{\mathrm{ren}}(u,s)  \right]  \,.
\end{equation}
It is then useful to define the signature-even combination 
\begin{equation}
L = -\ln(x) -  \frac{i\pi}{2}\approx \frac{1}{2}  \left( \ln\left( \Scale[0.8]{ \frac{-s-i \delta }{-t} } \right) + \ln\left( \Scale[0.8]{\frac{-u-i \delta }{-t} } \right)  \right)  
\end{equation}
and the color  operators~\cite{DelDuca:2013ara,DelDuca:2014cya}
\begin{align}
&\Scale[0.9]{  \mathbf{T}_s^2 = (\mathbf{T}_1 \!+\! \mathbf{T}_2)^a(\mathbf{T}_1\!+\! \mathbf{T}_2)^a ,  \;\quad \mathbf{T}_t^2 = (\mathbf{T}_1 \!+\! \mathbf{T}_3)^a(\mathbf{T}_1 \!+\! \mathbf{T}_3)^a,  }\nonumber \\
&\Scale[0.9]{ \mathbf{T}_u^2 = (\mathbf{T}_1 \!+\! \mathbf{T}_4)^a(\mathbf{T}_1 \!+\! \mathbf{T}_4)^a ,\;\quad \mathbf{T}_{s-u}^2 = \frac{1}{2}(\mathbf{T}_s^2 -\mathbf{T}_u^2 ).}
\end{align}

At leading power in $x$ and up to the next-to-leading logarithmic
(NLL) accuracy, i.e. up to terms of the form $\aso^i L^{i-1}$, the odd
amplitude has a simple factorized structure. Indeed, to all
orders in the strong coupling, $\mathcal H_{\rm ren,-}$ can
be thought of as the amplitude for the exchange of a single
``reggeized'' $t$-channel gluon, whose interaction with the external
high-energy gluons is described by so-called \emph{impact factors}~\cite{Lipatov:1976zz,Kuraev:1976ge,Fadin:1993wh,Collins:1977jy,Gribov:2009zz}.
In the language of complex angular momentum~\cite{Gribov:2003nw}, this
single-particle exchange is usually referred to as the ``Regge-pole''
contribution.

Starting from next-to-next-to-leading logarithmic (NNLL) accuracy
(i.e. from terms of the form $\aso^i L^{i-2}$), this simple
factorisation is broken and one needs to account for multiple Reggeons
exchanges~\cite{Gribov:2009zz,DelDuca:2001gu,DelDuca:2008pj,Caron-Huot:2013fea,Caron-Huot:2017fxr,Fadin:2016wso,Fadin:2017nka,Falcioni:2021buo}. These
are usually referred to as the ``Regge-cut'' contributions. For the
signature-even amplitude, the Regge cut contribution already enters at
the first non-trivial logarithmic order (NLL). The presence of Regge
cuts greatly increases the complexity of an all-order
analysis. However, if one restricts oneself to fixed order and only
considers the first non-trivial cut contribution (i.e. one works at
NLL/NNLL for the even/odd amplitude), the problem simplifies
dramatically. Indeed, this case can be dealt with using LO BFKL
theory~\cite{Gribov:2009zz,Caron-Huot:2013fea,Caron-Huot:2017fxr,Fadin:2016wso,Fadin:2017nka}.

The only
missing ingredient
to fully characterize the signature
even/odd amplitudes at NLL/NNLL and test Regge factorisation
to this accuracy 
is the three-loop gluon Regge trajectory. 
Currently, it is only known in $\mathcal N=4$ SYM~\cite{Henn:2016jdu,Falcioni:2021buo}, and in pure
gluodynamics under some assumptions on the trajectory itself~\cite{Jin:2019nya,DelDuca:2021vjq}.
The three-loop calculation presented in this letter allows us to
extract the trajectory in full QCD, closing this gap.

Before presenting our results, we note that the definition itself of a
Regge trajectory is subtle at NNLL~\cite{Caron-Huot:2017fxr,Fadin:2016wso,Fadin:2017nka,Falcioni:2021buo}. In this letter, for
definiteness we follow the ``Regge-cut'' scheme of ref.~\cite{Falcioni:2021buo}. In
particular, we write\footnote{In this section, we set the
renormalization scale to $\mu^2 = -t$.}
\begin{align}\label{eq:regge_L_expansion}
 \mathcal{H}_{\mathrm{ren},\pm}   &= \: Z_g^2  \:e^{L\mathbf{T}_t^2 \tau_g} \sum_{n=0}^3 \aso^n \sum_{k=0}^n L^k \mathbfcal{O}^{\pm,(n)}_k  \mathcal{H}_{\mathrm{ren}} ^{(0)}, 
\end{align}
where $\tau_g = \sum_{n=1} \aso^n \tau_n $ is the gluon Regge
trajectory and $Z_g = \sum_{n=0} \aso^n Z_g^{(n)}$ is a scalar factor
accounting for collinear singularities~\cite{Caron-Huot:2017fxr} whose explicit value is given in 
the Supplemental Material. The
non-vanishing odd signature color operators $\mathbfcal{O}^{-,(n)}_k
$ read up to NNLL~\cite{Caron-Huot:2017fxr}
\begin{align}
\label{eq:regge_odd_operators}
&  \mathbfcal{O}^{-,(0)}_0 =1,  \quad \mathbfcal{O}^{-,(1)}_0 
= 2\mathcal{I}^g_1,  \\
& \mathbfcal{O}^{-,(2)}_0 
= \left[ 2\mathcal{I}^g_2 +\left({\mathcal{I}^g_1}\right)^2\right] + \mathcal{B}^{-,\Scale[0.7]{(2)}} [ (\mathbf{T}^2_{s-u} )^2 - \Scale[1]{\frac{N_c^2}{4}}], \nonumber\\
& \mathbfcal{O}^{-,(3)}_1 =  \mathcal{B}_1^{-,\Scale[0.7]{(3)}}\mathbf{T}^2_{s-u}[\mathbf{T}^2_{t},\mathbf{T}^2_{s-u}] 
+ \mathcal{B}_2^{-,\Scale[0.7]{(3)}}[\mathbf{T}^2_{t},\mathbf{T}^2_{s-u}] \mathbf{T}^2_{s-u} ,  \nonumber
\end{align}
while the even signature ones are up to NLL~\cite{Caron-Huot:2017fxr}
% \begin{align}
% &   \mathbfcal{O}^{+,(1)}_0 = i \pi \,\mathcal{C}^{+,(1)} \, \mathbf{T}_{s-u}^2, \;  \mathbfcal{O}^{+,(2)}_1 = i \pi \, \mathcal{C}^{+,(2)}  \,[ \mathbf{T}_{t}^2, \mathbf{T}_{s-u}^2],  \nonumber \\
% & \mathbfcal{O}^{+,(3)}_2 = i \pi \, \mathcal{C}^{+,(3)} \, [\mathbf{T}_{t}^2,[ \mathbf{T}_{t}^2, \mathbf{T}_{s-u}^2]] . \label{eq:regge_even_operators} 
% \end{align}
\begin{align}
&   \mathbfcal{O}^{+,(1)}_0 = i \pi \,\mathcal{B}^{+,(1)} \, \mathbf{T}_{s-u}^2, \;  \mathbfcal{O}^{+,(2)}_1 = i \pi \, \mathcal{B}^{+,(2)}  \,[ \mathbf{T}_{t}^2, \mathbf{T}_{s-u}^2],  \nonumber \\
& \mathbfcal{O}^{+,(3)}_2 = i \pi \, \mathcal{B}^{+,(3)} \, [\mathbf{T}_{t}^2,[ \mathbf{T}_{t}^2, \mathbf{T}_{s-u}^2]] . \label{eq:regge_even_operators} 
\end{align}
In these equations, the coefficients $\mathcal{B}^{\pm,(L)}$ describe
the Regge cut contribution and are known~\cite{Caron-Huot:2013fea,Caron-Huot:2017fxr}. $\mathcal I^g_j$ are
the perturbative expansion coefficients of the gluon impact factor
and can be extracted from a one- and two-loop
calculation~\cite{Ahmed:2019qtg}.  For convenience, we report both $\mathcal{B}^{\pm,(L)}$ and $\mathcal I^g_{1,2}$ in the Supplemental Material.
As we noted earlier, the NNLL Regge trajectory instead requires a full
three-loop calculation. 
To present our result for it, we define
\begin{equation}\label{eq:kappa}
K(\as(\mu)) = - \frac{1}{4} \int_\infty^{\mu^2} \frac{d\lambda^2}{\lambda^2} \gamma^\text{K}\left(\as(\lambda^2)\right),
\end{equation}
together with its perturbative expansion $K = \sum_{n=1} K_i \aso^i$ whose coefficients are given in Supplemental Material. 
The expansion coefficients of the gluon Regge trajectory $\tau_i$
can then be written as
\begin{align}
\label{eq:regge_gluon_trajectory}
\tau_1 &= \; K_1 + \mathcal O(\ep),  \nonumber\\
\tau_2 &= \; K_2 
-\frac{56 n_f}{27}
+ N_c
\left( \frac{404}{27} - 2\zeta_3\right)
+ \mathcal O(\ep), \nonumber\\
\tau_3 &= \; K_3 +N_c^2 \bigg(16 \zeta_5 
+\frac{40 \zeta_2 \zeta_3}{3}-\frac{77 \zeta_4}{3}-\frac{6664 \zeta_3}{27} 
\nonumber\\
&-\frac{3196
   \zeta_2}{81}+\frac{297029}{1458}\bigg) 
+\frac{n_f}{N_c} \left(-4
   \zeta_4-\frac{76 \zeta_3}{9}+\frac{1711}{108}\right)
\nonumber\\
&+N_c n_f
   \left(\frac{412 \zeta_2}{81}+\frac{2 \zeta_4}{3}+\frac{632 \zeta_3}{9}-\frac{171449}{2916}\right) 
\nonumber\\
   &
   + n_f^2
   \left(\frac{928}{729}-\frac{128 \zeta_3}{27}\right)+
\mathcal{O}(\ep),
\end{align}
where the higher orders in $\epsilon$ for $\tau_1$ and $\tau_2$ can be found in the Supplemental
material.
As expected, our lower-loop results are consistent with ref.~\cite{Fadin:1996tb}, see also \cite{Blumlein:1998ib}.
For $\tau_3$, the $n_f$-independent part of our result agrees with
ref.~\cite{DelDuca:2021vjq}. Furthermore, the highest transcendental-weight terms of the trajectory agree with the $\mathcal N=4$ SYM result~\cite{Henn:2016jdu,Falcioni:2021buo},
as predicted by the maximal transcendentality principle~\cite{Kotikov:2001sc,Kotikov:2002ab,Kotikov:2004er,Kotikov:2007cy}.
On its own, the result~\eqref{eq:regge_gluon_trajectory} is not particularly illuminating. However, we have found the \emph{same}
trajectory using both the calculation outlined in this letter and our previous $qq'\to qq'$ three-loop calculation~\cite{Caola:2021rqz}.
This provides an important test of QCD Regge factorisation at the three-loop level. We also stress that now \emph{all} the
ingredients for a NLL/NNLL analysis of the signature-even/odd elastic amplitudes are known. In particular, we can now fully predict
the yet unknown $qg\to qg$ three-loop amplitude to NNLL accuracy. Explicitly checking these predictions against a full calculation
will provide a highly non-trivial test 
of the universality of Regge factorisation in QCD.

\section{Conclusion}
In this letter we have presented 
the first computation of the helicity amplitudes for the scattering 
of four gluons up to three loops in full QCD.  
We obtained compact results for the finite part of
all independent helicity configurations in terms of harmonic polylogarithms up to weight six and 
we  verified that the IR poles of our analytic amplitudes follow the 
predicted universal pattern up to three loops, which includes dipole and quadruple
correlations. We also considered the high-energy (Regge) limit of our amplitudes,
and extracted the full three-loop QCD gluon Regge trajectory. 
This was the last
missing building block to describe single-Reggeon exchanges at 
NNLL accuracy.
\newline

\acknowledgements
{\bf Acknowledgements.} 
We thank G.\ Falcioni, E.\ Gardi, N.\ Maher, C.\ Milloy, and
L.\ Vernazza  for discussions on the scheme of ref.~\cite{Falcioni:2021buo}, and for comparing unpublished results
for the three-loop gluon Regge trajectory and two-loop quark impact factors. The research of FC was supported by the ERC Starting Grant 804394 \textsc{hipQCD} and by the UK Science and Technology Facilities Council (STFC) under grant ST/T000864/1. GG was supported by the Royal Society grant URF/R1/191125. AvM was supported in part by the National Science Foundation through Grant 2013859. LT was supported by the Excellence Cluster ORIGINS funded by the Deutsche Forschungsgemeinschaft (DFG, German Research Foundation) under Germany’s Excellence Strategy - EXC-2094 - 390783311, by the ERC Starting Grant 949279 HighPHun and by the Royal Society grant URF/R1/191125.

\bibliographystyle{bibliostyle}
\bibliography{biblio}

\newpage
\onecolumngrid
\appendix

\section*{Supplemental material}
\makeatletter
\renewcommand\@biblabel[1]{[#1S]}
\makeatother

\subsection{UV renormalization}
In these sections we provide all the quantities whose explicitly values we omitted in the main text of this letter. \\
The $\beta$-function coefficients we used are  defined in the standard way through the following equations
\begin{equation}
\frac{d \as }{d \log \mu}  = \beta(\as) - 2\epsilon \as \; , \quad
\beta(\as) = -2 \as \sum\limits_{n=0} \beta_n \left(\frac{\as}{4 \pi}\right)^{n+1}  ,
\end{equation}
where  $\as\equiv\as(\mu)=4 \pi \aso $. We also recall the definition of standard $SU(N_c)$ Casimir constants: 
\begin{equation}
    C_A = N_c, \quad  C_F = \frac{N_c^2-1}{2N_c} \, .
\end{equation}
With this, up to third order of the perturbative expansion we have
\begin{align}
\beta_0 &= \frac{11}{3} C_A - \frac{2}{3}\: n_f \; , \nonumber\\
\beta_1 &= \frac{1}{3} \left(34 \: C_A^2-10\: C_A \:n_f \right)-2 \:C_F\: n_f \; ,\nonumber\\ 
\beta_2 &= -\frac{1415 \:C_A^2 \:n_f}{54}+\frac{2857\: C_A^3}{54}-\frac{205\: C_A\: C_F\:
   n_f}{18} +\frac{79 \:C_A \:n_f^2}{54}+C_F^2 \:n_f+\frac{11\: C_F \:n_f^2}{9} \;,
\end{align}
in terms of which the renormalized helicity amplitudes read
% \begin{align}\label{eq:hel_ampls_ren}
% \mathcal{H}_{\bm{\lambda},\text{ren}}^{(0)} &=  \mathcal{H}_{\bm{\lambda}}^{(0)} ,  \nonumber\\
% \mathcal{H}_{{\bm{\lambda}},\: \text{ren}}^{(1)} & = S_\epsilon^{-1}\mathcal{H}_{\bm{\lambda}}^{(1)}-\frac{\beta_0 }{\epsilon}    \mathcal{H}_{\bm{\lambda}}^{(0)},\nonumber\\ 
% \mathcal{H}_{{\bm{\lambda}},\: \text{ren}}^{(2)} &= S_\epsilon^{-2}\mathcal{H}_{\bm{\lambda}}^{(2)}  - \frac{2 \beta_0 }{\epsilon} S_\epsilon^{-1}\mathcal{H}_{\bm{\lambda}}^{(1)}  + \frac{ \left(2 \beta_0^2- \beta_1\epsilon\right)}{2 \epsilon^2} \mathcal{H}_{\bm{\lambda}}^{(0)}, \nonumber\\ 
% \mathcal{H}_{{\bm{\lambda}},\: \text{ren}}^{(3)} & = S_\epsilon^{-3}\mathcal{H}_{\bm{\lambda}}^{(3)} -\frac{3 \beta_0}{\epsilon}  S_\epsilon^{-2}\mathcal{H}_{\bm{\lambda}}^{(2)} +\frac{ \left(3 \beta_0^2-\beta_1 \epsilon\right)}{\epsilon^2} S_\epsilon^{-1}\mathcal{H}_{\bm{\lambda}}^{(1)} +\frac{ \left(7 \beta_1 \beta_0   \epsilon -6 \beta_0^3-2 \beta_2 \epsilon^2 \right)}{6 \epsilon^3} \mathcal{H}_{\bm{\lambda}}^{(0)}.  
% \end{align}
\begin{align}\label{eq:hel_ampls_ren}
\mathcal{H}_{\bm{\lambda},\text{ren}}^{(0)} &=  \mathcal{H}_{\bm{\lambda}}^{(0)} ,  \nonumber\\
\mathcal{H}_{{\bm{\lambda}},\: \text{ren}}^{(1)} & = \mathcal{H}_{\bm{\lambda}}^{(1)}-\frac{\beta_0 }{\epsilon}    \mathcal{H}_{\bm{\lambda}}^{(0)},\nonumber\\ 
\mathcal{H}_{{\bm{\lambda}},\: \text{ren}}^{(2)} &= \mathcal{H}_{\bm{\lambda}}^{(2)}  - \frac{2 \beta_0 }{\epsilon} \mathcal{H}_{\bm{\lambda}}^{(1)}  + \frac{ \left(2 \beta_0^2- \beta_1\epsilon\right)}{2 \epsilon^2} \mathcal{H}_{\bm{\lambda}}^{(0)}, \nonumber\\ 
\mathcal{H}_{{\bm{\lambda}},\: \text{ren}}^{(3)} & = \mathcal{H}_{\bm{\lambda}}^{(3)} -\frac{3 \beta_0}{\epsilon}  \mathcal{H}_{\bm{\lambda}}^{(2)} +\frac{ \left(3 \beta_0^2-\beta_1 \epsilon\right)}{\epsilon^2} \mathcal{H}_{\bm{\lambda}}^{(1)} +\frac{ \left(7 \beta_1 \beta_0   \epsilon -6 \beta_0^3-2 \beta_2 \epsilon^2 \right)}{6 \epsilon^3} \mathcal{H}_{\bm{\lambda}}^{(0)}.
\end{align}

\subsection{IR subtraction}
Similarly to the $\beta$-function coefficients, the cusp anomalous dimension and the gluon anomalous dimension admit a series expansion in $\as$:
\begin{equation}
\gamma^{\text{K}} = \sum\limits_{n=0} \left(\frac{\as}{4 \pi}\right)^{n+1}  \gamma_n^\text{K} ,  \qquad \gamma^g = \sum\limits_{n=0} \left(\frac{\as}{4 \pi}\right)^{n+1}  \gamma_n^g.
\end{equation}
The expansion coefficients of the cusp anomalous dimension read~\cite{Korchemsky:1987wg,Moch:2004pa,Vogt:2004mw,Grozin:2014hna}
\begin{align}
   \gamma_0^\text{K} &= 4 \,, \nonumber\\
   \gamma_1^\text{K} &= \left( \frac{268}{9} 
    - \frac{4\pi^2}{3} \right) C_A - \frac{40}{9}\, n_f \,,
  \nonumber\\
   \gamma_2^\text{K} &= C_A^2 \left( \frac{490}{3} 
    - \frac{536\pi^2}{27}
    + \frac{44\pi^4}{45} + \frac{88}{3}\,\zeta_3 \right) + C_A  n_f  \left(  \frac{80\pi^2}{27}- \frac{836}{27} - \frac{112}{3}\,\zeta_3 \right) + C_F n_f \left(32\zeta_3 - \frac{110}{3}\right) - \frac{16}{27}\, n_f^2 \; ,  
\end{align}
from which one can obtain the expansion of $K(\as)$ defined in eq.~\eqref{eq:kappa}:
\begin{align}
    K_1 &= \frac{\gamma_0^\text{K}}{\ep}\, , \nonumber\\
    K_2 &= \frac{2\gamma_1^\text{K}}{\ep} - \frac{\beta_0 \gamma_0^\text{K}}{2 \ep^2}\, ,\nonumber\\
    K_3 &= \frac{16\gamma_2^\text{K}}{3\ep} - \frac{4\beta_0 \gamma_1^\text{K} + 4\beta_1 \gamma_0^\text{K}}{3\ep^2} + \frac{\beta_0^2\gamma_0^\text{K}}{3\ep^3} \, .
\end{align}
The expansion coefficients of the gluon anomalous dimension in the notation of~\cite{Becher:2009qa} are \cite{Moch:2005id}
\begin{eqnarray}
   \gamma_0^g &=&   -\beta_0   \,, \nonumber\\
   \gamma_1^g  &=&    C_A^2 \left( -\frac{692}{27} + \frac{11}{3} \zeta_2 + 2 \zeta_3 \right)+C_A n_f \left( \frac{128}{27} - \frac{2}{3} \zeta_2 \right) + 2 C_F n_f \,,  \nonumber\\
   \gamma_2^g &=&   C_A^3\left(\frac{-97186}{729} + \frac{6109}{81} \zeta_2 + \frac{122}{3} \zeta_3- \frac{319}{3} \zeta_4 - \frac{40}{3} \zeta_2 \zeta_3\;- 16 \zeta_5 \right)+C_A^2 n_f \left( \frac{30715}{1458} - \frac{1198}{81}\zeta_2 + \frac{356}{27} \zeta_3 + \frac{82}{3} \zeta_4 \right)  \nonumber\\
&&\mbox{}+C_A C_F n_f \left( \frac{1217}{27} - 2 \zeta_2 - \frac{152}{9}\zeta_3 - 8 \zeta_4 \right) - C_F^2 n_f + C_A n_f^2 \left(-\frac{269}{1458} + \frac{20}{27} \zeta_2 - \frac{56}{27} \zeta_3 \right) - \frac{11}{9} C_F n_f^2.
\end{eqnarray}

In eq.~\eqref{eq:IR_factorisation} we introduced the IR regularization matrix $\mathbfcal{Z}_{IR} $,  which can also be expanded in the strong coupling as follows
\begin{equation}
\mathbfcal{Z}_{IR} = \sum_{n=0} \left( \frac{\as}{4 \pi}\right)^n \mathbfcal{Z}_n . 
\end{equation}
Here we give the coefficients appearing in the expansion above~\cite{Becher:2009qa, Caola:2021rqz}:
\begin{align}
 \mathbfcal{Z}_0 &=  1 \,,  \nonumber \\
 \mathbfcal{Z}_1 &=  \frac{\Gamma'_0}{4 \epsilon^2} + \frac{\mathbf{\Gamma}_0}{2 \epsilon} \,,   \label{eq:Zir}\nonumber\\
\mathbfcal{Z}_2 &=   \frac{{\Gamma_0'}^2}{32 \epsilon^4} + \frac{\Gamma'_0}{8 \epsilon^3} \left( \mathbf{\Gamma}_0 - \frac{3}{2} \beta_0  \right) +  \frac{\mathbf{\Gamma}_0}{8 \epsilon^2}(\mathbf{\Gamma}_0 - 2 \beta_0)  + \frac{\Gamma_1'}{16 \epsilon^2} + \frac{\mathbf{\Gamma}_1}{4 \epsilon}\,, \nonumber\\ 
\mathbfcal{Z}_3 &=  \frac{{\Gamma'_0}^3}{384 \epsilon^6}  + \frac{{\Gamma'_0}^2}{64 \epsilon^5}(\mathbf{\Gamma}_0 \!-\! 3 \beta_0) + \frac{\Gamma_0'}{32 \epsilon^4} \left( \mathbf{\Gamma}_0 \!- \!\frac{4}{3} \beta_0 \right) \left( \mathbf{\Gamma}_0 \!-\! \frac{11}{3} \beta_0 \right)   + \frac{\Gamma_0' \Gamma_1'}{64 \epsilon^4} +\frac{\mathbf{\Gamma}_0}{48\epsilon^3}(\mathbf{\Gamma}_0 - 2 \beta_0)(\mathbf{\Gamma}_0 - 4 \beta_0) \nonumber\\ 
&\quad +\! \frac{\Gamma'_0}{16 \epsilon^3} \left( \mathbf{\Gamma}_1\!- \!\frac{16}{9} \beta_1\right)  + \frac{\Gamma_1'}{32 \epsilon^3} \left( \mathbf{\Gamma}_0 \!-\! \frac{20}{9} \beta_0 \right)+ \frac{\mathbf{\Gamma}_0 \mathbf{\Gamma}_1}{8 \epsilon^2} - \frac{\beta_0 \mathbf{\Gamma}_1 + \beta_1 \mathbf{\Gamma}_0}{6 \epsilon^2} + \frac{\Gamma_2'}{36 \epsilon^2 } + \frac{\mathbf{\Gamma}_2 + \mathbf{\Delta}_4^{(3)}}{6 \epsilon} \; .
\end{align}
Above we have defined 
\begin{equation}
\Gamma'(\alpha_s) = \frac{\partial \mathbf{\Gamma}(\{p\},\alpha_s,\mu)}{\partial \log \mu} =  -\gamma^\text{K} \sum_i C_i  = \sum_{n=0} \left( \frac{\as}{4 \pi}\right)^{n+1} \Gamma_n' ,
\end{equation}
with the last equal sign giving the definition of the perturbative expansion coefficients $\Gamma_n' $.

From the infrared matrix $\mathbfcal{Z}_{IR}$ one can extract the scalar factor $Z_i =  \sum_{n=0} \left(\frac{\as}{4 \pi}\right)^n Z_i^{(n)},$  which for $i=g,q$ carries the collinear singularities of the external gluons or quarks respectively and was introduced in eq.~\eqref{eq:regge_L_expansion}. Its explicit expansion coefficients are
\begin{align}
 Z_i^{(0)} & = 1 \, , \nonumber\\
 Z_i^{(1)} & = -C_i \gamma^\text{K}_1 \frac{1}{ \ep^2} + 4\gamma_1^i\frac{1}{\ep} \, , \nonumber \\
 Z_i^{(2)} & = C_i^2  \frac{(\gamma^\text{K}_1)^2}{2 \ep^4} + C_i\left[\frac{1}{\ep^3}\gamma^\text{K}_1\left(\frac{3 \beta_0}{4} - 4\gamma_1^i \right) - \frac{\gamma^\text{K}_2}{\ep^2}  \right] + \frac{2}{\ep^2}\gamma_1^i\left(4\gamma_1^i -\beta_0 \right) + \frac{8\gamma_2^i}{ \ep}\, .
\end{align}
As a last ingredient we provide the analytical form of the quadrupole coefficient functions $D_1(x)$ and $D_2(x)$ in terms of harmonic polylogarithms with letters 0 and 1.  Explicitly~\cite{Almelid:2015jia,Henn:2020lye,Caola:2021rqz},
\begin{align}
D_1 &= -2 \textit{G}_{1,4}-\textit{G}_{2,3}-\textit{G}_{3,2}+2 \textit{G}_{1,1,3}+2 \textit{G}_{1,2,2}-2 \textit{G}_{1,3,0}-\textit{G}_{2,2,0}-\textit{G}_{3,1,0} +2 \textit{G}_{1,1,2,0}-2 \textit{G}_{1,2,0,0}+ 2 \textit{G}_{1,2,1,0}\nonumber\\
&\quad +4 \textit{G}_{1,0,0,0,0}-2 \textit{G}_{1,1,0,0,0}+\frac{\zeta_5}{2}  - 5 \zeta_2 \zeta_3 + \zeta_2[5 \textit{G}_{3}+5 \textit{G}_{2,0}+2 \textit{G}_{1,0,0}-6 (\textit{G}_{1,2}+\textit{G}_{1,1,0})]\nonumber\\
 &\quad + \zeta_3 (\textit{G}_{2}+2 \textit{G}_{1,0}-2 \textit{G}_{1,1}) 
 - i \pi  [-\zeta_3 \textit{G}_{0}+\textit{G}_{2,2}+\textit{G}_{3,0} +\textit{G}_{3,1}+ \textit{G}_{2,0,0}+2 (\textit{G}_{1,3}-\textit{G}_{1,1,2}-\textit{G}_{1,2,1}-\textit{G}_{1,0,0,0})]\nonumber\\
&\quad + i \pi \zeta_2 (-\textit{G}_{2}+2 (\textit{G}_{1,1}+\textit{G}_{1,0}))- 11i\pi \zeta_4 \, ,\label{D1}\\[10pt]
D_2 &=  2 \textit{G}_{2,3}+2 \textit{G}_{3,2}-\textit{G}_{1,1,3}-\textit{G}_{1,2,2}-2 \textit{G}_{2,1,2}+2 \textit{G}_{2,2,0}-2 \textit{G}_{2,2,1}  +2 \textit{G}_{3,1,0}-2 \textit{G}_{3,1,1}-\textit{G}_{1,1,2,0}- \textit{G}_{1,2,1,0}\nonumber\\
&\quad-2 \textit{G}_{2,1,1,0}+4 \textit{G}_{2,1,1,1}-\zeta_5 +4 \zeta_2 \zeta_3 + \zeta_3 \textit{G}_{1,1} +\zeta_2 [-6 \textit{G}_{3}-6 \textit{G}_{2,0}+2 \textit{G}_{2,1}+5 (\textit{G}_{1,2}+\textit{G}_{1,1,0})] \nonumber\\
&\quad + i \pi  (\zeta_3 \textit{G}_{1}+2 \textit{G}_{3,0}-\textit{G}_{1,1,2}-\textit{G}_{1,2,0}-\textit{G}_{1,2,1}+2 \textit{G}_{2,0,0}-2 \textit{G}_{2,1,0} +2 \textit{G}_{2,1,1}-\textit{G}_{1,1,0,0})+i \pi\zeta_2  (4 \textit{G}_{2}-\textit{G}_{1,1}) \, ,  \label{D2}
\end{align}
where the argument $x$ has been suppressed, and for the HPLs we used a compact notation similar to~\cite{Remiddi:1999ew,Maitre:2005uu}, 
\begin{align}
G_{a_1,\dots,a_n,\footnotesize\underbrace{  0,\dots,0}_{n_0}} = G(\underbrace{0,\dots,0}_{|a_1|-1},\text{sgn}(a_1),\dots,\underbrace{0,\dots,0}_{|a_n|-1},\text{sgn}(a_n),\underbrace{0,\dots,0}_{n_0};x). \nonumber
\end{align}
We provide the analytic expressions for these functions in the ancillary files attached to the arXiv submission of this paper.

\subsection{Impact factors, Regge cut coefficients and the Regge trajectory}
Below we provide all the constants appearing in eq.~\eqref{eq:regge_odd_operators}. We start from the expansion of the
gluon impact factor. 
At one loop we have
\begin{align}
\mathcal{I}^g_1 & =   N_c
   \left(4 \zeta_2-\frac{67}{18}\right)+\frac{5 n_f}{9}+\epsilon 
   \Bigg[N_c \left(\frac{17 \zeta_3}{3}+\frac{11 \zeta_2}{12}-\frac{202}{27}\right)+n_f \left(-\frac{\zeta_2}{6}+\frac{28}{27}\right)\Bigg]\nonumber\\
    &+\epsilon ^2 \Bigg[N_c \left(\frac{41
   \zeta_4}{8}+\frac{77 \zeta_3}{18}+\frac{67 \zeta_2}{36}-\frac{1214}{81}\right)+n_f \left(-\frac{7 \zeta_3}{9}-\frac{5 \zeta_2}{18}+\frac{164}{81}\right)\Bigg]\nonumber\\
    &+\epsilon ^3 \Bigg[N_c \left(-\frac{59 \zeta_2 \zeta_3}{6}+\frac{67 \zeta_5}{5}+\frac{517 \zeta_4}{96}+\frac{469 \zeta_3}{54}+\frac{101 \zeta_2}{27}-\frac{7288}{243}\right)+n_f \left(-\frac{47 \zeta_4}{48}-\frac{35 \zeta_3}{27}-\frac{14
   \zeta_2}{27}+\frac{976}{243}\right)\Bigg]\nonumber\\
   &+\epsilon ^4 \Bigg[N_c \left(-\frac{\pi ^6}{4320}-\frac{70 \zeta_3^2}{9}-\frac{77 \zeta_2 \zeta_3}{36}+\frac{341 \zeta_5}{30}+\frac{3149 \zeta_4}{288}+\frac{1414 \zeta_3}{81}+\frac{607 \zeta_2}{81}-\frac{43736}{729}\right)\nonumber\\
&+n_f \left(\frac{7 \zeta_2 \zeta_3}{18}-\frac{31 \zeta_5}{15}-\frac{235
   \zeta_4}{144}-\frac{196 \zeta_3}{81}-\frac{82 \zeta_2}{81}+\frac{5840}{729}\right)\Bigg] + \mathcal{O}(\ep^5)\,  \label{eq:impact_gluon_1}
\end{align}
and at two loops
\begin{align}
\mathcal{I}^g_2 & = -\frac{3 N_c^2 \zeta_2}{2 \epsilon ^2}+N_c^2 \left(\frac{9 \zeta_4}{4}+\frac{88 \zeta_3}{9}+\frac{335
   \zeta_2}{18}-\frac{26675}{648}\right)+N_c n_f
   \left(\frac{2 \zeta_3}{9}-\frac{25 \zeta_2}{9}+\frac{2063}{216}\right)+\frac{n_f}{N_c} \left(2 \zeta_3-\frac{55}{24}\right)-\frac{25
   n_f^2}{162}\nonumber\\
    &+\epsilon  \Bigg[N_c^2 \left(22 \zeta_2 \zeta_3-39 \zeta_5+\frac{275 \zeta_4}{4}+\frac{1865 \zeta_3}{18} +\frac{3191 \zeta_2}{72}-\frac{98671}{648}\right)+N_c n_f \left(-\frac{19 \zeta_4}{2}-\frac{157 \zeta_3}{9}-\frac{871
   \zeta_2}{108}+\frac{149033}{3888}\right)\nonumber\\
   &+\frac{n_f}{N_c} \left(3 \zeta_4+\frac{19 \zeta_3}{3}+\frac{\zeta_2}{4}-\frac{1711}{144}\right)+n_f^2 \left(\frac{5 \zeta_2}{54}-\frac{140}{243}\right)\Bigg] \nonumber\\
   &+\epsilon ^2 \Bigg[N_c^2 \left(-\frac{4733 \pi ^6}{30240}-\frac{659 \zeta_3^2}{18}-\frac{8987 \zeta_2 \zeta_3}{36}-\frac{187 \zeta_5}{5}+\frac{16103 \zeta_4}{64}+\frac{121859 \zeta_3}{324}+\frac{71263 \zeta_2}{648}-\frac{6140957}{11664}\right)\nonumber\\
   &+N_c n_f \left(\frac{781 \zeta_2 \zeta_3}{18}+\frac{104 \zeta_5}{5}-\frac{5803 \zeta_4}{144}-\frac{5698 \zeta_3}{81}-\frac{1645 \zeta_2}{72}+\frac{3197809}{23328}\right)\nonumber\\
   &+\frac{n_f}{N_c}
   \left(-2 \zeta_2 \zeta_3+14 \zeta_5+\frac{19 \zeta_4}{2}+\frac{197 \zeta_3}{9}+\frac{55 \zeta_2}{24}-\frac{42727}{864}\right)+n_f^2
   \left(-\frac{5 \zeta_4}{144}+\frac{35 \zeta_3}{81} + \frac{53 \zeta_2}{162}-\frac{404}{243}\right)\Bigg] + \mathcal{O}(\ep^3)\, . \label{eq:impact_gluon_2}
   \end{align}
   For the reader's convenience, we also report the quark impact factors
   at higher order in the dimensional regulator $\epsilon$, extracted from
   ref.~\cite{Caola:2021rqz}. 
   At one loop, we have
\begin{align}   
\mathcal{I}^q_1 & =\frac{4-\frac{
   \zeta_2}{2}}{N_c}+N_c \left(\frac{7
   \zeta_2}{2}+\frac{13}{18}\right)-\frac{5 n_f}{9}+\epsilon  \bigg[N_c
   \left(-\frac{\zeta_2}{6}+\frac{10 \zeta_3}{3}+\frac{40}{27}\right)+\frac{1}{N_c}\left(-\frac{3
   \zeta_2}{4}-\frac{7 \zeta_3}{3}+8\right)+n_f
   \left(\frac{\zeta_2}{6}-\frac{28}{27}\right)\bigg]\nonumber\\
   & +\epsilon ^2 \bigg[N_c \left(-\frac{13
   \zeta_2}{36}+\frac{35 \zeta_4}{16}-\frac{7 \zeta_3}{9}+\frac{242}{81}\right)+\frac{1}{N_c}\left(-2
   \zeta_2-\frac{47 \zeta_4}{16}-\frac{7 \zeta_3}{2}+16\right)+n_f \left(\frac{5
   \zeta_2}{18}+\frac{7 \zeta_3}{9}-\frac{164}{81}\right)\bigg]\nonumber\\
   & +\epsilon ^3 \bigg[N_c \left(-\frac{26 \zeta_2 \zeta_3}{3}-\frac{20 \zeta_2}{27}-\frac{47 \zeta_4}{48}+\frac{36
   \zeta_5}{5}-\frac{91 \zeta_3}{54}+\frac{1456}{243}\right)+\frac{1}{N_c}\left(\frac{7 \zeta_2 \zeta_3}{6}-4 \zeta_2-\frac{141
   \zeta_4}{32}-\frac{31 \zeta_5}{5}-\frac{28 \zeta_3}{3}+32\right)\nonumber\\
   &+n_f \left(\frac{14
   \zeta_2}{27}+\frac{47 \zeta_4}{48}+\frac{35 \zeta_3}{27}-\frac{976}{243}\right)\bigg]\nonumber\\
   & +\epsilon ^4 \bigg[\frac{1}{N_c}\left(\frac{7 \zeta_2 \zeta_3}{4}-8
   \zeta_2-\frac{47 \zeta_4}{4}-\frac{93 \zeta_5}{10}+\frac{49
   \zeta_3^2}{18}-\frac{56 \zeta_3}{3}-\frac{949 \pi
   ^6}{120960}+64\right) \nonumber\\
   &+N_c \left(\frac{7 \zeta_2 \zeta_3}{18}-\frac{121 \zeta_2}{81}-\frac{611 \zeta_4}{288}-\frac{31
   \zeta_5}{15}-\frac{91 \zeta_3^2}{18}-\frac{280 \zeta_3}{81}-\frac{977 \pi ^6}{120960}+\frac{8744}{729}\right)\nonumber\\
   &+n_f
   \left(-\frac{7 \zeta_2 \zeta_3}{18}+\frac{82
   \zeta_2}{81}+\frac{235 \zeta_4}{144}+\frac{31 \zeta_5}{15}+\frac{196 \zeta_3}{81}-\frac{5840}{729}\right)\bigg]  + \mathcal{O}(\ep^5)\, ,\label{eq:impact_quark_1}
\end{align}
      while at two loops
\begin{align}   
\mathcal{I}^q_2 & =-\frac{3 N_c^2
   \zeta_2}{2 \epsilon ^2} + N_c^2 \left(\frac{87 \zeta_2}{4}+\frac{25
   \zeta_4}{16}+\frac{41 \zeta
   _3}{9}+\frac{22537}{2592}\right)+\frac{1}{N_c^2}\left(\frac{21
   \zeta_2}{4}-\frac{83 \zeta_4}{16}-\frac{15 \zeta
   _3}{2}+\frac{255}{32}\right)\nonumber\\
   &+N_c n_f \left(-4
   \zeta_2-\frac{23 \zeta
   _3}{9}-\frac{650}{81}\right)+\frac{n_f}{N_c}
   \left(-\zeta_2-\frac{19 \zeta
   _3}{9}-\frac{505}{81}\right)+\frac{25
   n_f^2}{54}+\frac{19 \zeta_2}{2}-\frac{47
   \zeta_4}{8}-\frac{205 \zeta_3}{18}+\frac{28787}{648} \nonumber\\
&+\epsilon \bigg[ N_c^2 \left(\frac{161 \zeta_2 \zeta_3}{6}+\frac{4055
   \zeta_2}{144}+\frac{587 \zeta_4}{12}+\frac{49 \zeta
   _5}{2}+\frac{898 \zeta
   _3}{27}+\frac{911797}{15552}\right)+n_f^2
   \left(\frac{140}{81}-\frac{5 \zeta_2}{18}\right)\nonumber\\
   &+\frac{1}{N_c^2}\left(\frac{49 \zeta_2 \zeta
   _3}{6}+\frac{325 \zeta_2}{16}-\frac{201 \zeta_4}{16}-3 \zeta
   _5-\frac{166 \zeta_3}{3}+\frac{2157}{64}\right)+N_c
   n_f \left(-\frac{61 \zeta_2}{36}-\frac{247
   \zeta_4}{24}-\frac{85 \zeta
   _3}{27}-\frac{36031}{972}\right)\nonumber\\
   &+\frac{n_f}{N_c} \left(-\frac{13
   \zeta_2}{4}-\frac{83 \zeta_4}{24}-\frac{17 \zeta
   _3}{27}-\frac{11983}{486}\right)+13 \zeta_2
   \zeta_3+\frac{115 \zeta_2}{8}-\frac{1283
   \zeta_4}{48}+\frac{121 \zeta_5}{2}-\frac{5507 \zeta
   _3}{54}+\frac{746543}{3888} \bigg]\nonumber\\
   &+\epsilon^2 \bigg[N_c^2 \left(-\frac{3613 \zeta_2 \zeta_3}{18}+\frac{5131
   \zeta_2}{864}+\frac{31811 \zeta_4}{288}+\frac{94 \zeta
   _5}{5}-\frac{293 \zeta_3^2}{18}+\frac{12007 \zeta
   _3}{648}+\frac{3251 \pi
   ^6}{120960}+\frac{23246941}{93312}\right)\nonumber\\
   &+\frac{1}{N_c^2}\left(10 \zeta_2 \zeta
   _3+\frac{2287 \zeta_2}{32}-\frac{5627 \zeta_4}{64}-\frac{9
   \zeta_5}{2}+\frac{1255 \zeta_3^2}{18}-\frac{6205 \zeta
   _3}{24}+\frac{7193 \pi
   ^6}{120960}+\frac{13575}{128}\right)\nonumber\\
   &+N_c n_f
   \left(\frac{625 \zeta_2 \zeta_3}{18}+\frac{1475
   \zeta_2}{108}-\frac{779 \zeta_4}{72}-\frac{143 \zeta
   _5}{5}+\frac{1993 \zeta
   _3}{81}-\frac{805855}{5832}\right)\nonumber\\
   &+\frac{n_f}{N_c}\left(\frac{31
   \zeta_2 \zeta_3}{9}-\frac{45 \zeta_2}{4}-\frac{503
   \zeta_4}{144}-\frac{151 \zeta_5}{15}+\frac{623 \zeta
   _3}{81}-\frac{227023}{2916}\right)+n_f^2
   \left(-\frac{53 \zeta_2}{54}+\frac{5 \zeta_4}{48}-\frac{35 \zeta
   _3}{27}+\frac{404}{81}\right)\nonumber\\
   &+\frac{1613 \zeta_2 \zeta
   _3}{36}+\frac{197 \zeta_2}{24}-\frac{27175
   \zeta_4}{144}+\frac{791 \zeta_5}{30}+\frac{1621 \zeta
   _3^2}{18}-\frac{170951 \zeta_3}{324}+\frac{17 \pi
   ^6}{70}+\frac{16114247}{23328} \bigg]+ \mathcal{O}(\ep^3) \, .\label{eq:impact_quark_2}
\end{align}
These results are also provided in electronic format in the \texttt{arXiv} submission of this manuscript.\\
Moving to the cut coefficients we have for the odd signature ones \cite{Caron-Huot:2017fxr,Falcioni:2021buo}
\begin{align}
\mathcal{B}^{-,(2)}& = \frac{2\pi^2}{3} r_\Gamma^2 \left( \frac{3}{\ep^2} - 18 \ep \zeta_3 - 27 \ep^2 \zeta_4+ \mathcal{O}(\ep) \right),\nonumber \\
\mathcal{B}_1^{-,(3)} & = 64\pi^2 r_\Gamma^3  \left( \frac{1}{48\ep^2} + \frac{37}{24} \zeta_3 + \mathcal{O}(\ep) \right) ,\nonumber \\
\mathcal{B}_2^{-,(3)} &= 64\pi^2 r_\Gamma^3  \left( \frac{1}{24\ep^2} + \frac{1}{12} \zeta_3 + \mathcal{O}(\ep) \right).
\end{align}
while for even signature
\begin{align}
&   \mathcal{B}^{+,(1)} = \, r_{\Gamma} \: \frac{2}{\epsilon},  \nonumber\\
& \mathcal{B}^{+,(2)} = - \frac{r_{\Gamma}^2}{2} \left(  \frac{4}{\epsilon^2} +
 72 \zeta_3 \epsilon + 108 \zeta_4 \epsilon^2 +\mathcal{O}(\epsilon^3)\right),  \nonumber \\
& \mathcal{B}^{+,(3)} = \frac{r_{\Gamma}^3}{6} \bigg(  \frac{8}{\epsilon^3}  - 176 \zeta_3 - 264 \zeta_4 \epsilon  -  5712 \zeta_5 \epsilon^2  + \mathcal{O}(\epsilon^3) \bigg) . 
\end{align}
Finally, we report the gluon Regge trajectory at higher order in the dimensional regulator $\ep$. At one loop we have the exact result
\begin{align}
\tau_1 &= \; e^{\ep \gamma_E}  \frac{\Gamma(1-\ep)^2 \Gamma(1+\ep)}
{\Gamma(1-2\ep)} \frac{2}{\ep},
\end{align}
while at two loops
\begin{align}
\tau_2 &= \; -\frac{\beta_0}{\ep^2} + \frac{1}{\epsilon}\left[ -\frac{10 n_f}{9} + N_c
\left( \frac{67}{9} - 2\zeta_2 \right) \right]  -\frac{56 n_f}{27} 
 + N_c
\left( \frac{404}{27} - 2\zeta_3\right) +\epsilon \bigg[n_f
   \left(12 \zeta_3 -\frac{328}{81}+\frac{5
   \pi ^2}{27}\right) \nonumber \\
&+ N_c \left(\frac{2428}{81}-66 \zeta_3-\frac{67 \zeta_2}{9}-3\zeta_4\right) \bigg]+\epsilon^2 \bigg[N_c \bigg(82\zeta_5
+\frac{142 \zeta_2 \zeta_3}{3}-\frac{4556 \zeta_3 }{27}+\frac{14576}{243}-\frac{404 \zeta_2}{27}-\frac{2321 \zeta_4}{24}\bigg)\nonumber \\
&+n_f \left(\frac{680
   \zeta_3}{27}-\frac{1952}{243}+\frac{56 \zeta_2}{27}+\frac{211 \zeta_4}{12}\right)\bigg] + \mathcal{O}(\ep^3).
\end{align}

\end{document}